\documentclass[aps, amsmath,, a4paper, floatfix]{revtex4}

\usepackage{graphicx}
\usepackage{amsmath}
\usepackage{amsfonts}
\usepackage{amssymb}
\usepackage{color}
\usepackage{dcolumn}

\usepackage{bm}% bold math

\usepackage{hyperref}% add hypertext capabilities

\begin{document}
\title{Weak-localization and spin-orbit interaction in side-gate field effect devices at the LaAlO$_3$/SrTiO$_3$ interface}

\author{D. Stornaiuolo$^{1,2}$, S. Gariglio$^1$, A. F\^ete$^{1}$, M. Gabay$^3$, D. Li$^1$, D. Massarotti$^2$ and J.-M. Triscone$^{1}$}

\affiliation{$^{1}$DQMP - University of Geneva, 24 quai Ernest-Ansermet CH-1211 Geneva, Switzerland\\$^{2}$Department of Physics, University of Naples Federico II and CNR-SPIN Napoli, Italy \\$^{3}$Laboratoire de Physique des Solides, Bat 510, Universit{\'e} Paris-Sud, 91405 Orsay Cedex, France}

\date{\today}

\begin{abstract}
Using field effect devices with side gates, we modulate the 2 dimensional electron gas hosted at the LaAlO$_3$/SrTiO$_3$ interface to study the temperature and doping evolution of the magnetotransport. The analysis of the data reveals different transport regimes depending on the interplay between the different (elastic, inelastic, and spin-orbit) scattering times and their temperature dependencies. We find that the spin-orbit interaction strongly affects the low temperature transport in the normal state in a very large region of the phase diagram, extending beyond the superconducting dome.
\end{abstract}
\maketitle

\section{Introduction}

Spin-orbit (SO) coupling offers a way to manipulate the spin degree of freedom of the charge carriers through the control of their motion \cite{winkler}. The realization of devices where such control is achieved is at the core of spintronics and can have an enormous impact on applications ranging from memories to quantum computing. Recently, spin-orbit interaction has received great interest also in relation with superconductivity, due to the possibility to induce unconventional superconducting order parameters \cite{gorkov,cappelluti}, and likewise in the search for Majorana states in hybrid structures,  where a semiconductor nanowire, hosting SO coupling, is placed in contact with superconducting electrodes \cite{mourik,das,rokhinson,alicea,beenakker}.\\
The 2-dimensional electron gas (2DEG) present at the LaAlO$_3$/SrTiO$_3$ (LAO/STO) interface exhibits Rashba SO coupling which is tunable using the electric field effect \cite{caviglia_prlso,fete_PRB,shalom, fete_NJP}.
Differently from 2DEG in semiconductor heterostructures, the carriers at the LAO/STO interface are hosted by 3\textit{d} electronic orbitals, potentially impacting the strength and nature of the SO coupling \cite{zhong}.
Moreover, LAO/STO interfaces undergo a superconducting transition, also widely tunable by electric gating \cite{caviglia_nature08}.

The aim of this work is to identify the various transport regimes of the 2DEG across a wide range of
carrier concentrations and temperatures, in order to map the relation between SO and superconductivity in this system. To this purpose, we realized nanoscale field effect devices with a side-gate configuration. This new layout allowed us to perform a local modulation of the carrier concentration at the LAO/STO interface with unprecedented efficiency and to reach very low carrier densities while keeping good metallic contacts for the voltage probes.

We used these devices to study the magnetotransport characteristics of LAO/STO and we uncover a crossover line between a transport regime described by weak-localisation (WL) and one characterized by weak-antilocalisation (WAL), where spin-orbit interaction is the dominant scattering mechanism. We find that the temperature range where WAL is dominant, initially restricted to the millikelvin values for the strongly depleted state, becomes wider and wider when increasing the carriers concentration. Moreover, we point out that
the transition to the superconducting phase always takes place from within the Rashba SO coupling dominated transport regime.

\section{Devices design and fabrication}

\begin{figure}[b]
\centering
\includegraphics[width=4.5 cm, height=9.5 cm]{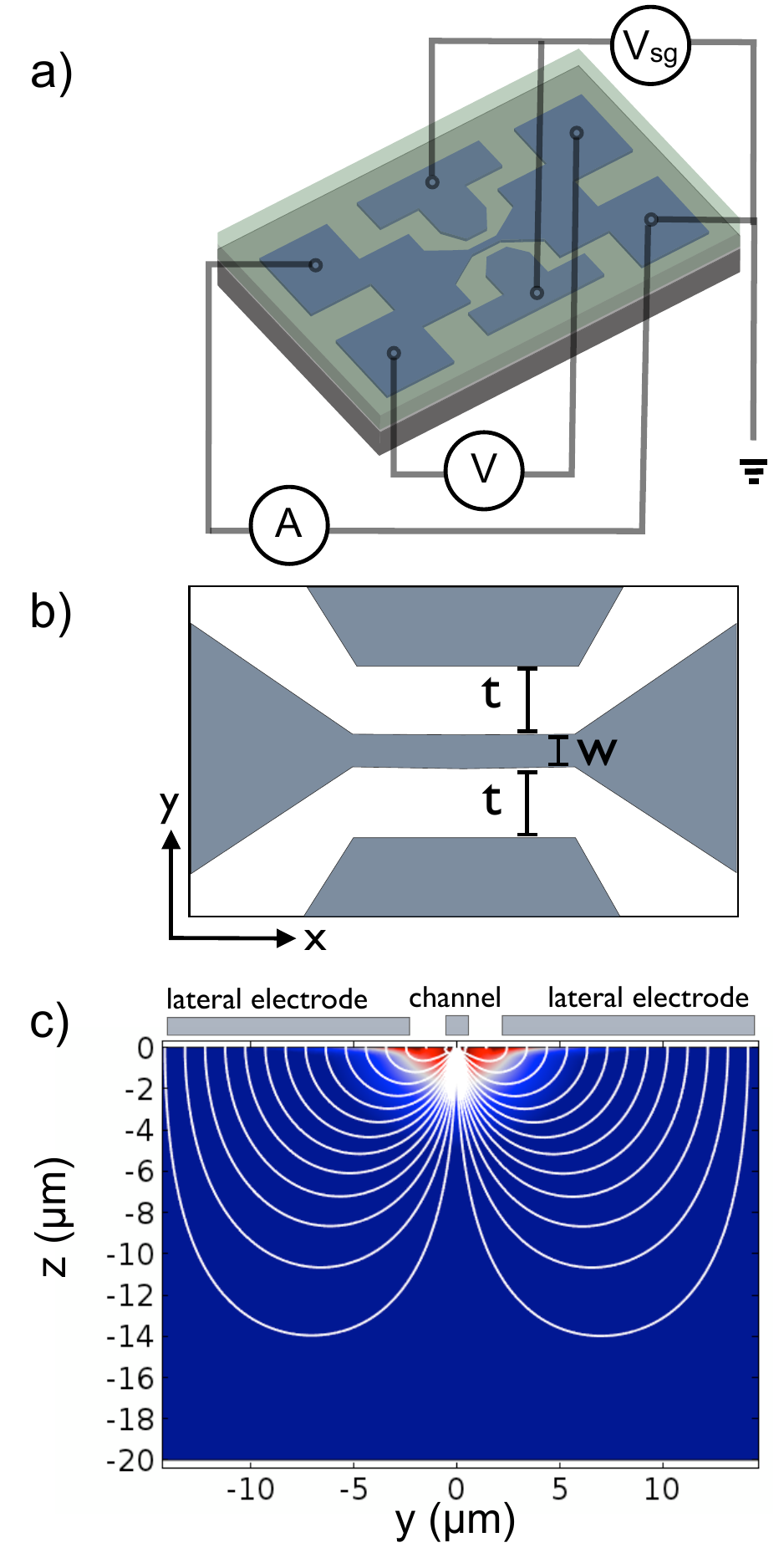}
\caption{(Color online) (a) Sketch of the device geometry showing the electrical connections used to realize measurements under field effect. The blue area indicates the 2DEG. (b) Zoom of the central area of the device: $w$ is the channel width and $t$ is the distance between each side gate and the channel edge. (c) Finite element simulations of the electric field intensity: the $yz$ plane at the middle of the channel is shown. The electric field lines are drawn in white.}
\label{fig1}
\end{figure}

The field effect has been successfully used to uncover many of the exceptional properties of the LAO/STO system \cite{caviglia_nature08,bell,caviglia_prlso}. Most of the work realized up to now uses a back-gate configuration, with an electrode deposited on the back of the STO single crystal substrate. This configuration capitalizes on the high dielectric constant of STO, although it requires large gate voltages due to the thickness of the gate dielectric, typically 0.5~mm.
More recently, top gate configurations were also developed, using LAO as the gate dielectric \cite{forg, hosoda} or realizing electric double layer transistors \cite{lin}. 
In the present work, we use a side-gate configuration, as sketched in Figure \ref{fig1}: this set-up allows us to take advantage of the high dielectric constant of STO, using a reduced effective thickness of the dielectric gate and thereby decreasing the applied gate voltages \cite{pellegrino_thesis}.\\

\begin{figure}[t]
\centering
\includegraphics[width=8.0 cm, height=5.5 cm]{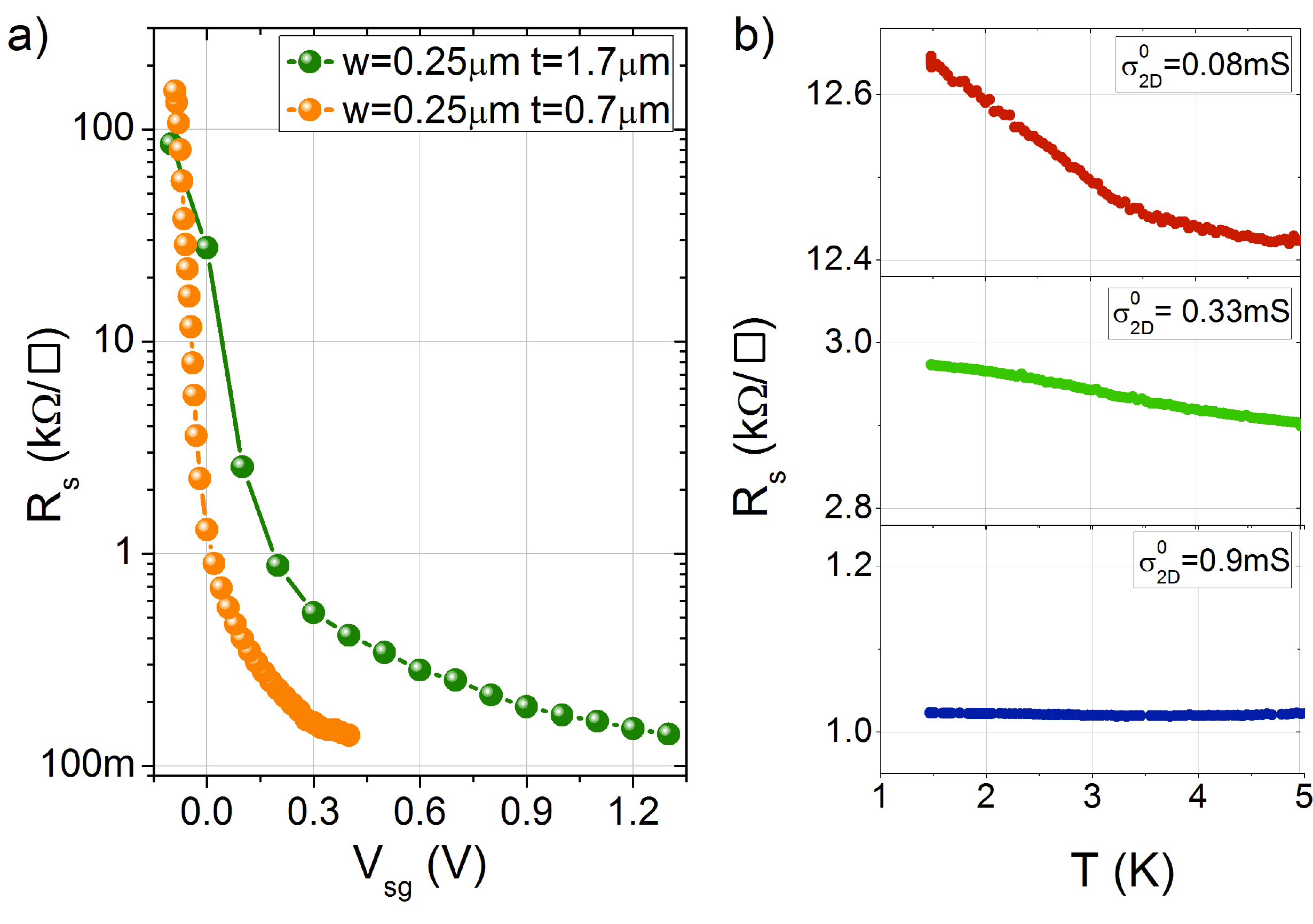}
\caption{(Color online) (a) Sheet resistance $R_s$ vs. gate voltage $V_{sg}$ for two side-gate field effect devices having the same width $w$=0.25~$\mu$m and different distance $t$ from the side electrodes to the channel edge ($t$=1.7~$\mu$m for green data and 0.7~$\mu$m for orange data). (b) $R_s$ vs. temperature T for three selected carrier concentration. The carrier concentration is indicated using the value of the sheet conductance $\sigma_{2D}^0$, as explained in the text.}
\label{fig2}
\end{figure}

In order to design efficient devices, we performed finite element simulations of the side gate configuration using the Comsol Multiphysics software. Figure \ref{fig1}(c) shows a cut of the device along the $yz$ plane, $x$ being the axis oriented along the channel and in the plane of the conduction. The thickness of the section in the $z$ direction is 20~$\mu$m. As can be seen, the electric field lines (drawn in white) depart from the side electrodes and penetrate into the STO before bending and closing in correspondence with the 2DEG channel. The effect on the gas is thus comparable to a back-gate action, with the electric field lines reaching the gas from below. The confinement of the electric field inside the STO substrate is due to its high dielectric permittivity \cite{pellegrino}. The simulations were done assuming, for the low temperature STO dielectric constant $\epsilon$, the electric field $E$ dependence: \begin{math} \epsilon(E) =1+ \frac{B}{[1+(E/E_0)^2]^{1/3}} \end{math} with $B$=25462 and $E_0$=82213~Vm$^{-1}$\cite{sakudo,neville,matthey}. The simulation reveals a focusing effect of the field lines on the conducting path. In this way, a modulation of the carrier density of 10$^{14}$cm$^{-2}$~V$^{-1}$ can be obtained for a 0.5~$\mu$m wide channel separated from the lateral electrodes by 1~$\mu$m; \textit{i.e.} a 1~V gate sweep can completely deplete the channel (see the Supplemental Material for calculations of the charge induced in side gate devices of various dimensions.). We point out that, using the same simulation tool, we calculated a charge modulation of only 10$^{11}$cm$^{-2}$~V$^{-1}$ for the conventional back-gate geometry, assuming a channel of several tens of microns. Increasing the channel width $w$ or the separating dielectric gap $t$ strongly reduces the charge modulation (see the Supplementary Material). For this reason, the channels studied in this work have width $w$ ranging from 0.25 to 1.3~$\mu$m, gaps $t$ from 0.7 to 1.7~$\mu$m and length $L$ between 2 and 4~$\mu$m.\\
In order to fabricate these nanoscale devices, we followed the e-beam based patterning technique described in Ref.~\onlinecite{stornaiuolo}. We realize a templated substrate where an amorphous STO layer delimits the areas where the channel, the bonding pads and the lateral gate electrodes will be created. Afterwards, we deposit by pulsed laser deposition \cite{cancellieri} a 10 unit cell film of LAO which will grow epitaxially only in the channel, pads and gates. As the gate electrodes are realized simultaneously with the 2DEG channel, this technique does not require any further fabrication step, and ensures minimal manipulation of the sample.\\
Figure \ref{fig2}(a) shows a comparison of the modulation of the sheet resistance $R_s$, measured at $T$=1.5~K, as a function of the applied gate voltage $V_{sg}$ of two devices having the same width $w$ and different dielectric gaps $t$: the voltage efficiency is higher for the device having $t$=0.7~$\mu$m than for the $t$=1.7~$\mu$m one, as expected from the simulations. For the first device, a gate voltage sweep of 0.6~V changes $R_s$ by three orders of magnitude. We also note that close to the insulating divergence of the resistance, the conductance swing is two orders of magnitude for $\Delta V_{sg}\sim$90~mV. During all the measurements, the leakage current remained below the nA, even for the smallest gap. As a comparison, field effect devices with a conventional back-gate configuration require several hundred volts to perform a tuning of the same amplitude \cite{caviglia_nature08}. Moreover, using the side-gate structures we can extend our studies to extremely low doping levels, almost one order of magnitude below what is usually attained using the back-gate configuration. This is due to better focusing of the field lines on the channel, an effect that results in a local modulation of the carrier density; the voltage probes, unaffected by the gate voltage, are hence not charge depleted and provide a good metallic contact throughout the experiment.
\\
As expected, the application of the gate voltage changes the transport behavior of the 2DEG dramatically. Panel (b) of Figure \ref{fig2} shows the $R_s$ vs. temperature behavior for three levels of the field modulated carrier concentration. The carrier concentration is indicated using the value of the sheet conductance $\sigma_{2D}^0$, as the two quantities have a one to one correspondence \cite{fete_PRB,sigmavsn2D}. Reducing the carrier concentration, going from higher to lower values of  $\sigma_{2D}^0$, the sheet resistance shows a more and more pronounced increase at low temperature, indicating the emergence of an insulating ground state.

\section{Temperature and doping evolution of the magnetoconductance curves}

We measured the magnetoresistance of side-gate field effect devices for various carrier concentrations and, for each of these, at various temperatures.
Measurements in the temperature range form 1.5~K to 10~K were performed in a flow cryostat, whereas measurements in the temperature range from 0.3~K to 1.5~K were performed in a Heliox system, equipped with RC and copper powder filters to ensure good thermalization and low noise measurements. We also present measurements performed in a dilution cryostat with a base temperature of 50~mK.\\

\begin{figure}[t]
\centering
\includegraphics[width=7.5 cm, height=4.5 cm]{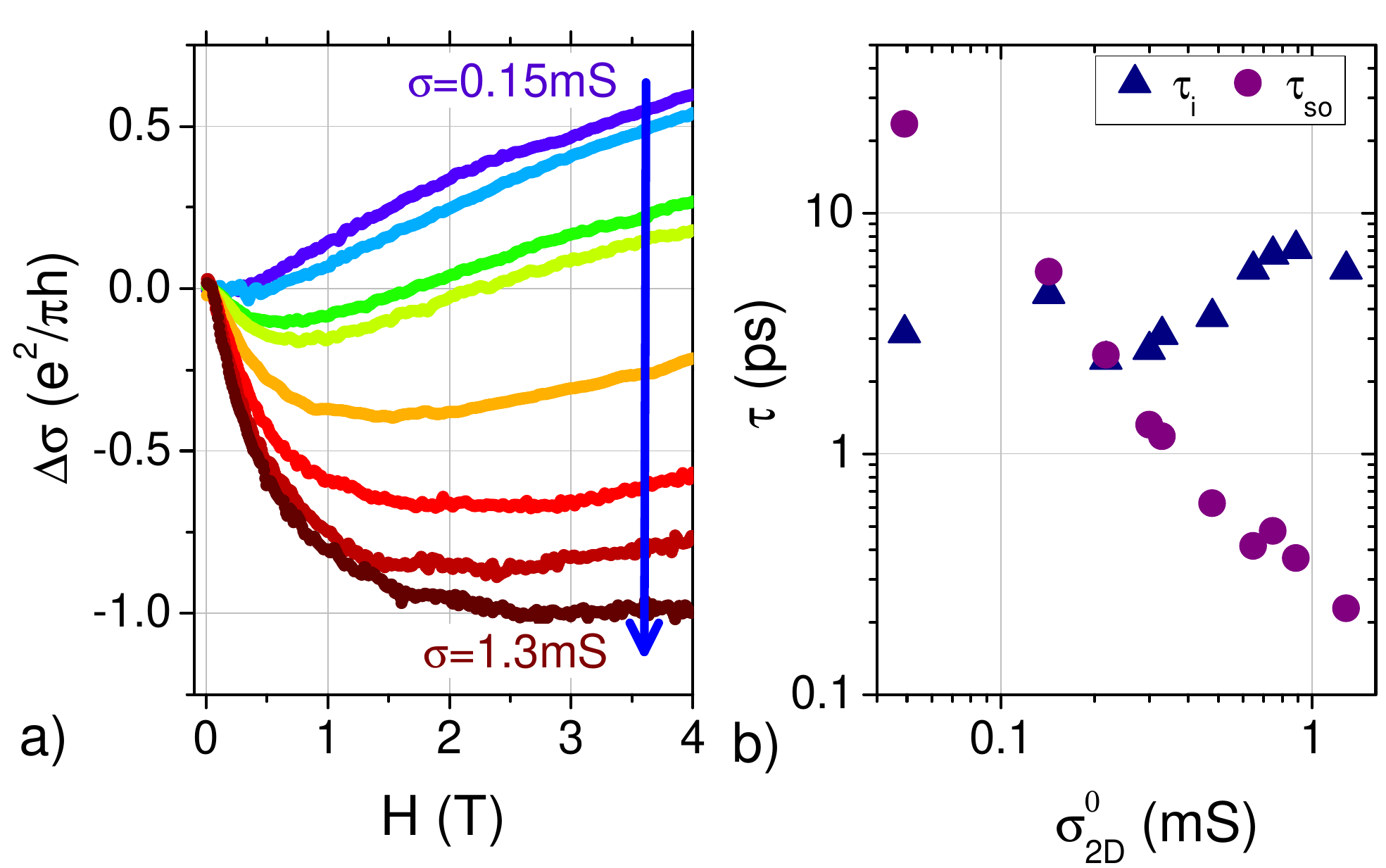}
\caption{(Color online) Magnetoconductance curves of a side-gate field effect device with $w$=0.75~$\mu$m, $t$=1.7~$\mu$m and $L$=4~$\mu$m measured at $T$=1.5~K. Panel (a) shows the modulation of the magnetoconductance curves varying the sheet conductance $\sigma_{2D}^0$ from 0.15~mS (top curve) to 1.3~mS (bottom curve). Panel (b) shows the inelastic ($\tau_i$) and spin-orbit ($\tau_{so}$) scattering times extracted from the curves.}
\label{fig3}
\end{figure}

We first show that the transport behavior of these micro-channel side-gate devices is consistent with the one observed in large (500~$\mu$m wide) path samples in the conventional back-gate configuration. Figure~\ref{fig3} displays the magnetoconductance measurements of a side-gate field effect device ($w$=0.75~$\mu$m, $t$=1.7~$\mu$m and $L$=4.5~$\mu$m) at $T$=1.5~K for different values of the sheet conductance at zero magnetic field ($\sigma_{2D}^0$). Also in this case, the sheet conductance is taken as a reference for the field effect modulated carrier concentration.
In figure~\ref{fig3} $\Delta\sigma =\sigma_{2D}(H)-\sigma_{2D}^0$ is plotted in units of $e^2/\pi h$, with the magnetic field $H$ applied perpendicular to the interface. 
The curves in this figure are in quantitative agreement with results reported in Ref.~\onlinecite{fete_PRB}. 

We fitted the magnetoconductance curves shown in Figure~\ref{fig3} using the Maekawa-Fukuyama (MF) formula \cite{maekawa}, following Ref.~\onlinecite{caviglia_prlso}:

\begin{eqnarray}\label{MF}
    \frac{\Delta\sigma(H)}{\sigma_0}=\Psi\left( \frac{H}{H_i +H_{so}}\right)+\nonumber
    \\
    +\frac{1}{2\sqrt{1-\gamma^2}}\Psi\left( \frac{H}{H_i +H_{so}(1+\sqrt{1-\gamma^2})}\right)+\nonumber
    \\
    -\frac{1}{2\sqrt{1-\gamma^2}}\Psi\left( \frac{H}{H_i +H_{so}(1-\sqrt{1-\gamma^2})}\right)
\end{eqnarray}

where $\Psi(x)= ln(x)+\psi(\frac{1}{2}+\frac{1}{x})$  (with $\psi(x)$ the digamma function), $\sigma_0=e^2/\pi h$,  and $\gamma=g\mu_B H/4eDH_{so}$ (with $\mu_B$ the Bohr magneton and $e$ the electron charge). The diffusion constant $D$ is given by the Drude expression: 
\begin{equation}
D=\frac{1}{2} \upsilon_{F}^{2} \tau_{el}
\label{D}
\end{equation}
where $\upsilon_{F}=\hslash k_F/m^*$ (with $k_F=\sqrt{2\pi n_{2D}}$ the Fermi momentum and $n_{2D}$ the number of carriers). The scattering time $\tau_{el}$ is given by: $\tau_{el}=m^*\mu/e$ where $\mu=\sigma_{2D}^0/e n_{2D}$. The estimation of the carrier density $n_{2D}$ and of the effective mass $m^*$ is rather difficult as the system evolves from one band $d_{xy}$ conduction to a two-band $d_{xy}$ and $d_{xz}$/$d_{yz}$ conduction when the filling is changed with the gate voltage \cite{joshua2012universal} \cite{fete_PRB}. 
For the current analysis, we have taken $n_{2D}$ as the total number of carriers calculated from the high magnetic field value of the Hall resistance \cite{Hall}, while for the effective mass $m^*$ we refer to estimation presented in Ref. \cite{fete_PRB} from the analysis of the WL correction. Although this analysis prevents the tracking of the evolution of the scattering times of each band, Rainer and Bergmann \cite{PhysRevB.32.3522} have shown that the MF formula remains valid also in presence of a complex Fermi surface. The spin-orbit and inelastic times extracted using this procedure are therefore values averaged over the two-bands; they provide nonetheless the evolution of the conductance  of the 2DEG for different dopings and temperatures.

From the fitting we extracted the inelastic $H_i$ and the spin-orbit $H_{so}$ characteristic fields and the effective $g$ factor included in the Zeeman correction $\gamma$. We calculated the inelastic and spin-orbit scattering times $\tau_i$ and $\tau_{so}$, respectively, using the relation: $H_{i,so}=\hbar/4eD\tau_{i,so}$. \\
As shown in panel (b) of Figure~\ref{fig3}, we observe that at 1.5~K the crossover from WAL to WL occurs at $\sigma_{2D}^0\approx$0.2~mS, in agreement with our previous measurements \cite{fete_PRB}.
This analysis confirms that side-gate devices behave as standard LAO/STO interfaces tuned by back-gate voltages. \\

\begin{figure}[t]
\centering
\includegraphics[width=8.5 cm, height=7.5 cm]{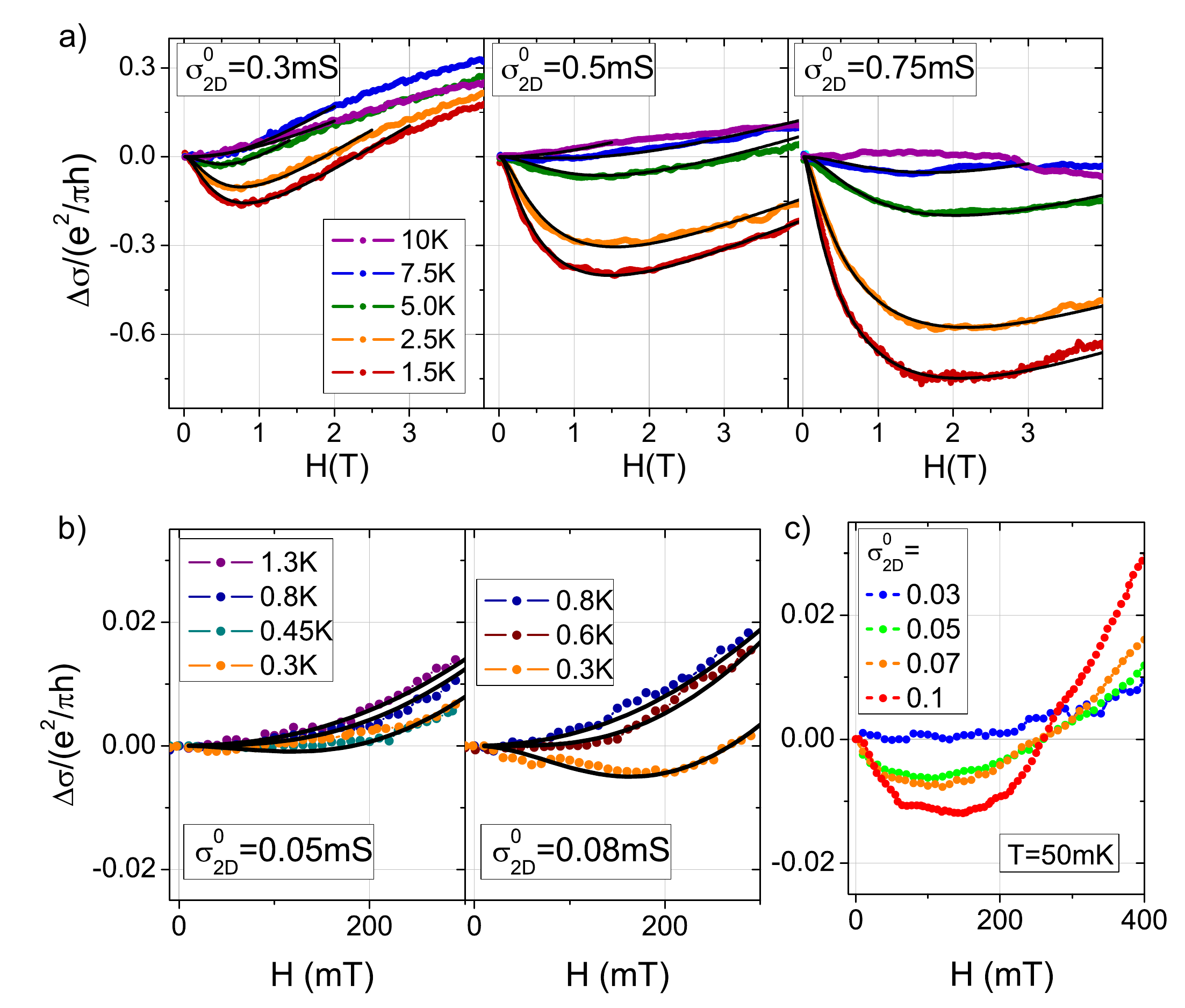}
\caption{(Color online) Magnetoconductance curves as a function of the temperature for selected doping levels, identified by the sheet conductance $\sigma_{2D}^0$. Panel (a) shows curves acquired in a flow cryostat in the temperature range from 1.5~K to 10~K, panel (b) shows curves acquired in a Heliox system with a base temperature of 0.3~K, and panel (c) shows curves acquired in a dilution cryostat keeping the temperature at 50~mK and changing the carrier concentration. Note the different field scales. Data in panels (a) and (b) refer to a 0.75~$\mu$m wide channel with $t$=1.7~$\mu$m and $L$=4.5~$\mu$m, whereas data in panel (c) refer to a  0.5~$\mu$m wide channel with $t$=1.7~$\mu$m and $L$=4.5~$\mu$m. The black solid lines in panels (a) and (b) are the fits performed using Eq.\ref{MF}.}
\label{fig4}
\end{figure}

We now turn to the evolution of the magnetoconductance as a function of temperature. Panels (a) and (b) of Figure~\ref{fig4} show magnetoconductance curves of side-gate devices for selected sheet conductances at various temperatures (also in this case, we show only the  curves for positive fields).  For each $\sigma_{2D}^0$, the magnetoconductance at low fields is negative, at least at the lowest temperature (red curves in panel (a), corresponding to T=1.5~K, and orange curves in panel (b), corresponding to T=0.3~K). Upon increasing the temperature, the minimum of the magnetoconductance curve is shifted upward, until the values become positive at high temperature. For the highest conductance shown ($\sigma_{2D}^0$=0.75~mS ), we observe a direct transition from a WAL behavior to a standard quadratic magnetoconductance. This happens because, for increasing conductance, along with the decrease in the spin-orbit scattering time, we observe an increase in the elastic scattering time \cite{fete_PRB}, hence suppressing the WL regime.

\begin{figure}[b]
\centering
\includegraphics[width=6.0 cm, height=7 cm]{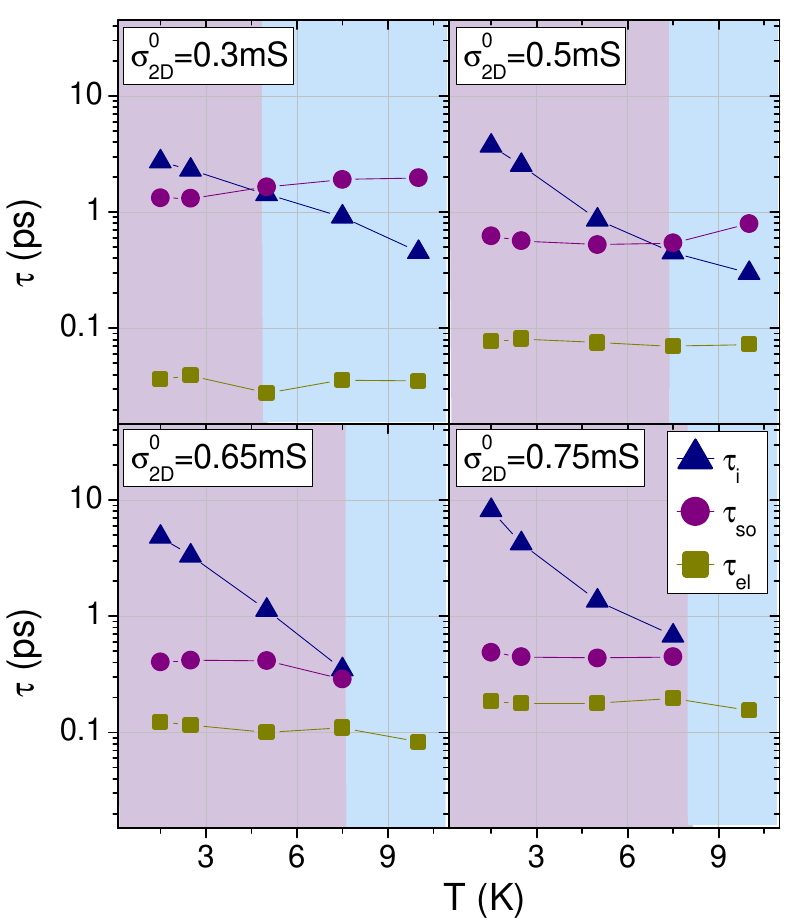}
\caption{(Color online) Evolution of inelastic ($\tau_i$), spin orbit ($\tau_{so}$) and elastic ($\tau_{el}$) scattering times for the 0.75~$\mu$m wide channel as a function of the temperature for selected values of the sheet conductance. In the pink region spin-orbit interaction is the main scattering mechanism with $\tau_i>\tau_{so}$. The crossing point between $\tau_i$ and $\tau_{so}$ identifies the crossover temperature $T_{cross}$ between WL and WAL.}
\label{fig5}
\end{figure}

Panel (c) shows magnetoconductance curves measured in a dilution cryostat at T=50~mK for very low carriers concentrations, where superconductivity is suppressed. In this case, we tuned the gate voltage of the 2DEG keeping the temperature constant in order  to avoid thermalization problems at these low temperatures. The WAL/WL transition takes place for a sheet conductance $\sigma_{2D}^0$ between 0.05~mS and 0.03~mS.
Fits to the data using equation 1 are shown as black solid lines in Figure~\ref{fig4}. The scattering times $\tau_i$, $\tau_{so}$ and $\tau_{el}$ extracted from the fits are shown in Figure~\ref{fig5} for selected sheet conductances.

\begin{figure}[t]
\centering
\includegraphics[width=8.5 cm, height=7 cm]{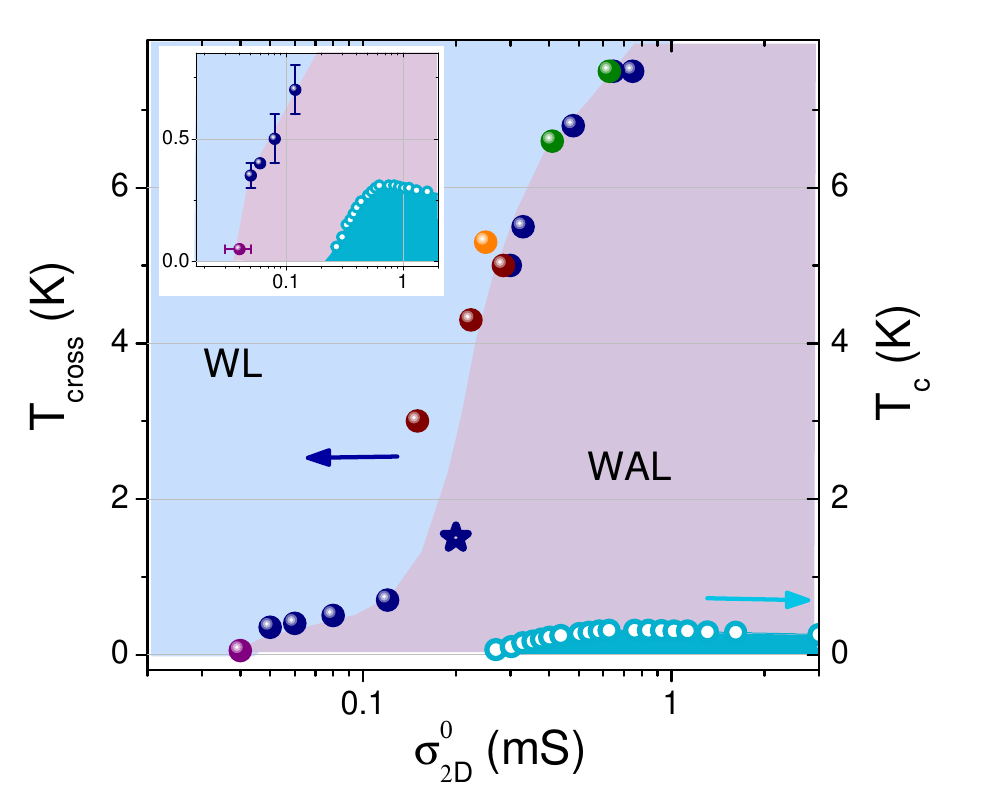}
\caption{(Color online) Phase diagram showing the temperature and doping for which LAO/STO magnetotransport is characterized by WAL or WL. The points represent the temperatures $T_{cross}$ for which $\tau_i=\tau_{so}$. The different colors of the data points refer to side-gate field effect devices of different size: $w$=1.3~$\mu$m, $t$=1.7~$\mu$m, $L$=8~$\mu$m (green dots), $w$=0.75~$\mu$m  $t$=1.7~$\mu$m, $L$=4.5~$\mu$m (blue dots), $w$=0.5~$\mu$m, $t$=1.7~$\mu$m,  $L$=4.5~$\mu$m (purple dot), $w$=0.25~$\mu$m, $t$=0.7~$\mu$m  $L$=2.5~$\mu$m (orange dot). The blue star is the crossing point extracted from the data of Figure \ref{fig3}, in agreement with the results obtained in Refs.~\onlinecite{caviglia_prlso} and \onlinecite{fete_PRB}. The brown dots refer to a sample with $w$=500~$\mu$m  $L$=1~mm in the conventional back-gate configuration. The blue area at the bottom indicates the superconducting dome (right axis for the superconducting $T_c$). The inset is a zoom of the low doping region. For $\sigma_{2D}^0$=0.05~mS we obtained $T_{cross}$=50~mK (see panel c of Figure 4).}
\label{fig6}
\end{figure}

The analysis reveals a temperature independent $\tau_{so}$ while $\tau_i$ displays a power law increase with decreasing temperature. We define a crossover temperature $T_{cross}$ between WAL and WL as the temperature at which $\tau_i=\tau_{so}$, signaling a change in the dominant scattering mechanism.
$T_{cross}$ as a function of $\sigma_{2D}^0$ is reported in the phase diagram of Figure~\ref{fig6}. Data were collected on several side-gate samples with channel width going from 0.25 to 1.3~$\mu$m (see Figure caption). We also include data from a large scale channel, 500$\mu$m wide, measured using the conventional back-gate configuration (brown points). 
The pink (light blue) region labelled WAL (WL) in Figure 6 is the area where $\tau_i>\tau_{so}$ ($\tau_i<\tau_{so}$). 
The light blue circles indicate the superconducting transition temperature $T_c$ (right axis) measured on different samples. This superconducting phase diagram is indeed a reproducible feature of LAO/STO devices. 
The phase diagram of Figure~\ref{fig6} shows that, also at extremely low carrier concentration, the magnetoconductance curves show a negative region, indicating WAL, albeit in a reduced range of temperatures \cite{Hsosmall}.

\section{Discussion}

A remarkable feature of the data presented in Figure \ref{fig6} is the sharp increase of $T_{cross}$ at $\sigma_{2D}^0\sim 0.2$~mS. A non-linear behavior of the spin orbit characteristic field as a function of the carrier concentration was already observed \cite{caviglia_prlso, fete_PRB}, and a correspondence was found between the sharp increase in the spin orbit strength, extracted from magnetoconductance curves measured at 1.5~K, and the emergence of superconductivity. BenShalom and coworkers \cite{shalom} have extracted the spin orbit coupling energy from the superconducting properties; also in this case, the SO interaction was found to follow the evolution of the superconducting transition temperature. This is in agreement with the observations that the superconducting critical field for the configuration parallel to the interface exceeds the paramagnetic limit \cite{shalom,reyren_apl}. This behavior can be related to the influence of a strong SO coupling, relaxing the Clogston limit of the depairing field for singlet superconductivity\cite{clogston,wu}.\\
The data we present in Figure \ref{fig6} are in agreement with these results.
The evolution of SO coupling with carrier concentration can be analyzed taking into account the evolution of the band filling in LAO/STO. In this system, the confinement of the conducting electrons at the interface induces a splitting in the Ti 3d-$t_{2g}$ orbitals, with the $d_{xy}$ bands having the lowest energy\cite{salluzzo}. At very low carrier concentrations (in the depletion regime), only the $d_{xy}$ bands are filled. Using field effect, the carrier concentration can be increased and the $d_{xz}$/$d_{yz}$ bands starts to be populated. We point out that the two band scenario is also in agreement with the analysis of Shubnikov de Haas oscillations reported in several works \cite{shalom2010shubnikov, caviglia2010two}. The progressive band filling has been shown to be related to the sharp enhancement of the Rashba field $H_{so}$\cite{fete_PRB}, and explains the extension of the temperature region where SO is the dominant scattering mechanism (Figure \ref{fig6}).
The data at very low sheet conductance ($\sigma_{2D}^0<$0.1~mS) in Figure 6 refer to a transport regime where the 2DEG does not show superconductivity. According to the above filling scenario, the transport in this range would be dominated by the $d_{xy}$ bands, with WAL clearly visible only in the millikelvin temperature range.
 
\section{Conclusions}

In conclusion, by realizing side-gate field effect devices, we were able to study the magnetoconductance of the 2DEG hosted at the LAO/STO interface in a wide range of carrier concentration and as a function of the temperature, down to the millikelvin range. 
The analysis of the temperature dependence of the scattering times extracted from the magnetoconductance fits confirms the weak localization scenario. The data indicate the presence of WAL due to SO interaction also at extremely low $\sigma_{2D}^0$.  Mapping the strength of the SO coupling, we observe that the transition to the superconducting state always takes place from a WAL regime. The side-gate approach opens the way to study in detail the depletion state of the LAO/STO interface and potentially can lead to field effect devices with extremely steep subthreshold swings.

\section{Acknowledgements}
We are deeply grateful for the fruitful discussions with A.F. Morpurgo and to
N.J.G. Couto for his assistance with the electron beam lithography. We also thank Jennifer Fowlie for careful reading of the manuscript and Marco Lopes and
S{\'e}bastien Muller for their technical help. The research leading to these results has received funding from the European Research Council under the European Union’s Seventh Framework Programme (FP7/2007-2013) / ERC Grant Agreement n.319286.  This work was also supported by the Swiss National
Science Foundation (Switzerland) through the National Center of Competence in Research, Materials with Novel
Electronic Properties (MaNEP) and Division II, by the Italian MIUR FIRB project HybridNanoDev RBFR1236VV001 and by the Institut Universitaire de France .

\clearpage

\section*{\boldmath S\lowercase{upplementary material - }W\lowercase{eak-localization and spin-orbit interaction in side-gate field effect devices at the }L\lowercase{a}A\lowercase{l}O$_3$/S\lowercase{r}T\lowercase{i}O$_3$\lowercase{ interface}}

\setcounter{figure}{0}

\subsection*{Induced charge in the side gate configuration}

We calculated the charge induced in side gate field effect devices using finite element simulations (see main text for details).
Panel (a) of Figure 1S shows the calculated induced charge density per Volt considering a channel of width $w$=0.5~$\mu$m and varying distance $t$ of the lateral electrodes (see the sketch in the right part of the panel). In panel (b) the calculations were performed keeping fixed the distance $t$ and varying $w$.\\ These simulations show how increasing the channel width $w$ or the separating gap $t$ reduces strongly the charge modulation in this side gate configuration. In order to obtain efficient devices, we realized samples where both the channel width $w$ and the distance $t$ are smaller than 2$\mu$m. 
\\
\\
\\
\begin{figure}[h]
\centering
\includegraphics[width=9 cm, height=7.5 cm]{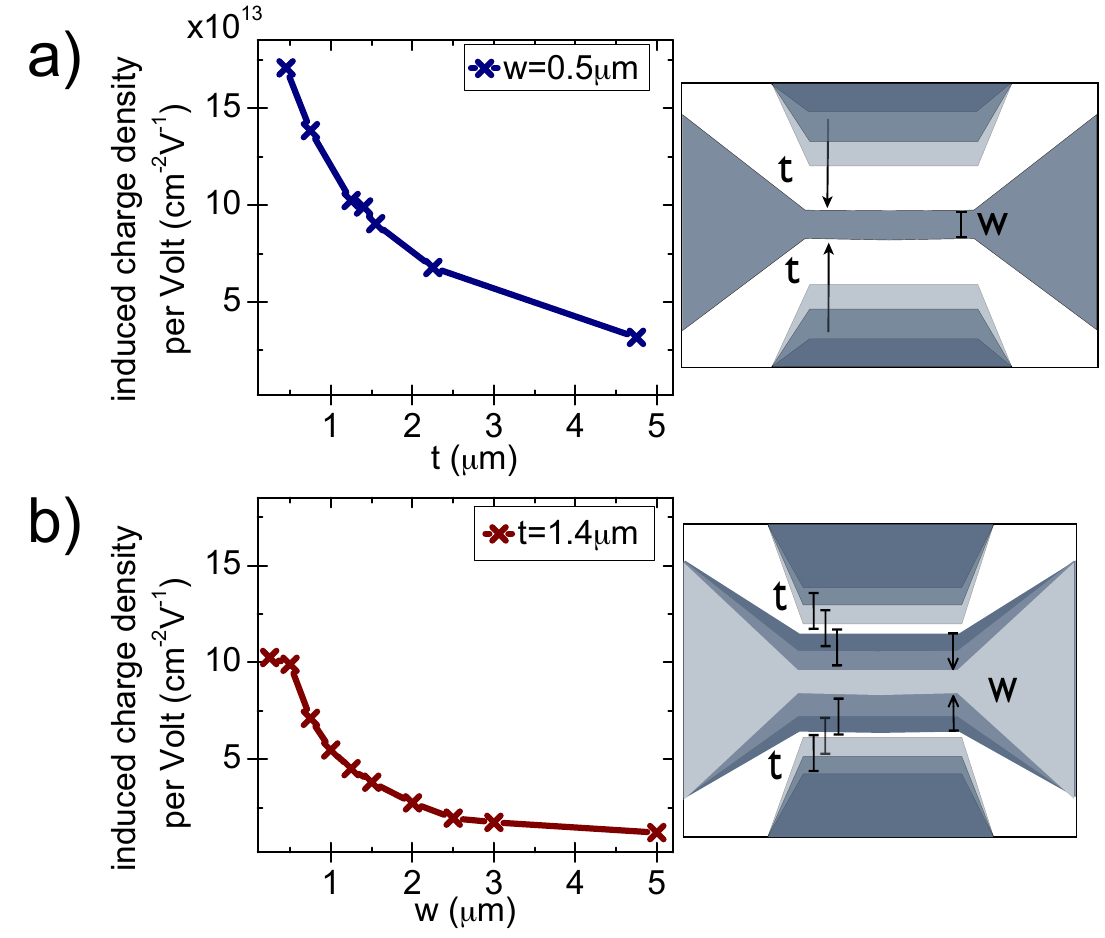}
\caption{(Color online) Panel (a) shows the induced charge density per Volt for a channel with $w$=0.5~$\mu$m calculated  as a function of the distances $t$ between the channel edge and the side gate electrodes, as shown by the arrows in the sketch in the right part. Panel (b) shows the induced charge density per Volt for a device with $t$=1.4~$\mu$m calculated as a function of the channel width $w$ (arrows in the right part of the panel).}
\label{fig1S}
\end{figure}


\begin{thebibliography}{35}
\expandafter\ifx\csname natexlab\endcsname\relax\def\natexlab#1{#1}\fi
\expandafter\ifx\csname bibnamefont\endcsname\relax
  \def\bibnamefont#1{#1}\fi
\expandafter\ifx\csname bibfnamefont\endcsname\relax
  \def\bibfnamefont#1{#1}\fi
\expandafter\ifx\csname citenamefont\endcsname\relax
  \def\citenamefont#1{#1}\fi
\expandafter\ifx\csname url\endcsname\relax
  \def\url#1{\texttt{#1}}\fi
\expandafter\ifx\csname urlprefix\endcsname\relax\def\urlprefix{URL }\fi
\providecommand{\bibinfo}[2]{#2}
\providecommand{\eprint}[2][]{\url{#2}}

\bibitem[{\citenamefont{Winkler}(2003)}]{winkler}
\bibinfo{author}{\bibfnamefont{R.}~\bibnamefont{Winkler}},
  \emph{\bibinfo{title}{Spin-orbit coupling effects in two-dimensional electron
  and hole systems}} (\bibinfo{publisher}{Springer}, \bibinfo{year}{2003}).

\bibitem[{\citenamefont{Gor'kov and Rashba}(2001)}]{gorkov}
\bibinfo{author}{\bibfnamefont{L.~P.} \bibnamefont{Gor'kov}} \bibnamefont{and}
  \bibinfo{author}{\bibfnamefont{E.~I.} \bibnamefont{Rashba}},
  \bibinfo{journal}{Phys. Rev. Lett.} \textbf{\bibinfo{volume}{87}},
  \bibinfo{pages}{037004} (\bibinfo{year}{2001}).

\bibitem[{\citenamefont{Cappelluti et~al.}(2007)\citenamefont{Cappelluti,
  Grimaldi, and Marsiglio}}]{cappelluti}
\bibinfo{author}{\bibfnamefont{E.}~\bibnamefont{Cappelluti}},
  \bibinfo{author}{\bibfnamefont{C.}~\bibnamefont{Grimaldi}}, \bibnamefont{and}
  \bibinfo{author}{\bibfnamefont{F.}~\bibnamefont{Marsiglio}},
  \bibinfo{journal}{Phys. Rev. Lett.} \textbf{\bibinfo{volume}{98}},
  \bibinfo{pages}{167002} (\bibinfo{year}{2007}).

\bibitem[{\citenamefont{Mourik et~al.}(2012)\citenamefont{Mourik, Zuo, Frolov,
  Plissard, Bakkers, and Kouwenhoven}}]{mourik}
\bibinfo{author}{\bibfnamefont{V.}~\bibnamefont{Mourik}},
  \bibinfo{author}{\bibfnamefont{K.}~\bibnamefont{Zuo}},
  \bibinfo{author}{\bibfnamefont{S.}~\bibnamefont{Frolov}},
  \bibinfo{author}{\bibfnamefont{S.}~\bibnamefont{Plissard}},
  \bibinfo{author}{\bibfnamefont{E.}~\bibnamefont{Bakkers}}, \bibnamefont{and}
  \bibinfo{author}{\bibfnamefont{L.}~\bibnamefont{Kouwenhoven}},
  \bibinfo{journal}{Science} \textbf{\bibinfo{volume}{336}},
  \bibinfo{pages}{1003} (\bibinfo{year}{2012}).

\bibitem[{\citenamefont{Das et~al.}(2012)\citenamefont{Das, Ronen, Most, Oreg,
  Heiblum, and Shtrikman}}]{das}
\bibinfo{author}{\bibfnamefont{A.}~\bibnamefont{Das}},
  \bibinfo{author}{\bibfnamefont{Y.}~\bibnamefont{Ronen}},
  \bibinfo{author}{\bibfnamefont{Y.}~\bibnamefont{Most}},
  \bibinfo{author}{\bibfnamefont{Y.}~\bibnamefont{Oreg}},
  \bibinfo{author}{\bibfnamefont{M.}~\bibnamefont{Heiblum}}, \bibnamefont{and}
  \bibinfo{author}{\bibfnamefont{H.}~\bibnamefont{Shtrikman}},
  \bibinfo{journal}{Nat. Phys.} \textbf{\bibinfo{volume}{8}},
  \bibinfo{pages}{887} (\bibinfo{year}{2012}).

\bibitem[{\citenamefont{Rokhinson et~al.}(2012)\citenamefont{Rokhinson, Liu,
  and Furdyna}}]{rokhinson}
\bibinfo{author}{\bibfnamefont{L.~P.} \bibnamefont{Rokhinson}},
  \bibinfo{author}{\bibfnamefont{X.}~\bibnamefont{Liu}}, \bibnamefont{and}
  \bibinfo{author}{\bibfnamefont{J.~K.} \bibnamefont{Furdyna}},
  \bibinfo{journal}{Nat. Phys.} \textbf{\bibinfo{volume}{8}},
  \bibinfo{pages}{795} (\bibinfo{year}{2012}).

\bibitem[{\citenamefont{Alicea}(2012)}]{alicea}
\bibinfo{author}{\bibfnamefont{J.}~\bibnamefont{Alicea}},
  \bibinfo{journal}{Rep. Prog. Phys.} \textbf{\bibinfo{volume}{75}},
  \bibinfo{pages}{076501} (\bibinfo{year}{2012}).

\bibitem[{\citenamefont{Beenakker}(2013)}]{beenakker}
\bibinfo{author}{\bibfnamefont{C.}~\bibnamefont{Beenakker}},
  \bibinfo{journal}{Ann. Rev. Cond. Matter Phys.} \textbf{\bibinfo{volume}{4}},
  \bibinfo{pages}{113} (\bibinfo{year}{2013}).

\bibitem[{\citenamefont{Caviglia
  et~al.}(2010{\natexlab{a}})\citenamefont{Caviglia, Gabay, Gariglio, Reyren,
  Cancellieri, and Triscone}}]{caviglia_prlso}
\bibinfo{author}{\bibfnamefont{A.~D.} \bibnamefont{Caviglia}},
  \bibinfo{author}{\bibfnamefont{M.}~\bibnamefont{Gabay}},
  \bibinfo{author}{\bibfnamefont{S.}~\bibnamefont{Gariglio}},
  \bibinfo{author}{\bibfnamefont{N.}~\bibnamefont{Reyren}},
  \bibinfo{author}{\bibfnamefont{C.}~\bibnamefont{Cancellieri}},
  \bibnamefont{and} \bibinfo{author}{\bibfnamefont{J.-M.}
  \bibnamefont{Triscone}}, \bibinfo{journal}{Phys. Rev. Lett.}
  \textbf{\bibinfo{volume}{104}}, \bibinfo{pages}{126803}
  (\bibinfo{year}{2010}{\natexlab{a}}).

\bibitem[{\citenamefont{F{\^e}te et~al.}(2012)\citenamefont{F{\^e}te, Gariglio,
  Caviglia, Triscone, and Gabay}}]{fete_PRB}
\bibinfo{author}{\bibfnamefont{A.}~\bibnamefont{F{\^e}te}},
  \bibinfo{author}{\bibfnamefont{S.}~\bibnamefont{Gariglio}},
  \bibinfo{author}{\bibfnamefont{A.D}~\bibnamefont{Caviglia}},
  \bibinfo{author}{\bibfnamefont{J.-M.} \bibnamefont{Triscone}},
  \bibnamefont{and} \bibinfo{author}{\bibfnamefont{M.}~\bibnamefont{Gabay}},
  \bibinfo{journal}{Phys. Rev. B} \textbf{\bibinfo{volume}{86}},
  \bibinfo{pages}{201105} (\bibinfo{year}{2012}).

\bibitem[{\citenamefont{BenShalom et~al.}(2010{\natexlab{a}})\citenamefont{BenShalom,
  Sachs, Rakhmilevitch, Palevski, and Dagan}}]{shalom}
\bibinfo{author}{\bibfnamefont{M.}~\bibnamefont{ BenShalom }},
  \bibinfo{author}{\bibfnamefont{M.}~\bibnamefont{Sachs}},
  \bibinfo{author}{\bibfnamefont{D.}~\bibnamefont{Rakhmilevitch}},
  \bibinfo{author}{\bibfnamefont{A.}~\bibnamefont{Palevski}}, \bibnamefont{and}
  \bibinfo{author}{\bibfnamefont{Y.}~\bibnamefont{Dagan}},
  \bibinfo{journal}{Phys. Rev. Lett.} \textbf{\bibinfo{volume}{104}},
  \bibinfo{pages}{126802} (\bibinfo{year}{2010}{\natexlab{a}}).
  
\bibitem[{\citenamefont{fete et~al.}(2014{\natexlab{a}})\citenamefont{F{\^e}te,
  Gariglio, Berthod, Li, Stornaiuolo, Gabay, and Triscone}}]{fete_NJP}
\bibinfo{author}{\bibfnamefont{A.}~\bibnamefont{ F{\^e}te}},
  \bibinfo{author}{\bibfnamefont{S.}~\bibnamefont{Gariglio}},
  \bibinfo{author}{\bibfnamefont{C.}~\bibnamefont{Berthod}},
  \bibinfo{author}{\bibfnamefont{D.}~\bibnamefont{Li}},
  \bibinfo{author}{\bibfnamefont{D.}~\bibnamefont{Stornaiuolo}},
  \bibinfo{author}{\bibfnamefont{M.}~\bibnamefont{Gabay}}, \bibnamefont{and}
  \bibinfo{author}{\bibfnamefont{J.-M.}~\bibnamefont{Triscone}},
  \bibinfo{journal}{New Journal of Physics} \textbf{\bibinfo{volume}{16}},
  \bibinfo{pages}{112002} (\bibinfo{year}{2014}{\natexlab{a}}). 

\bibitem[{\citenamefont{Zhong et~al.}(2013)\citenamefont{Zhong, T{\'o}th, and
  Held}}]{zhong}
\bibinfo{author}{\bibfnamefont{Z.}~\bibnamefont{Zhong}},
  \bibinfo{author}{\bibfnamefont{A.}~\bibnamefont{T{\'o}th}}, \bibnamefont{and}
  \bibinfo{author}{\bibfnamefont{K.}~\bibnamefont{Held}},
  \bibinfo{journal}{Phys. Rev. B} \textbf{\bibinfo{volume}{87}},
  \bibinfo{pages}{161102} (\bibinfo{year}{2013}).

\bibitem[{\citenamefont{Caviglia et~al.}(2008)\citenamefont{Caviglia, Gariglio,
  Reyren, Jaccard, Schneider, Gabay, Thiel, Hammerl, Mannhart, and
  Triscone}}]{caviglia_nature08}
\bibinfo{author}{\bibfnamefont{A.}~\bibnamefont{Caviglia}},
  \bibinfo{author}{\bibfnamefont{S.}~\bibnamefont{Gariglio}},
  \bibinfo{author}{\bibfnamefont{N.}~\bibnamefont{Reyren}},
  \bibinfo{author}{\bibfnamefont{D.}~\bibnamefont{Jaccard}},
  \bibinfo{author}{\bibfnamefont{T.}~\bibnamefont{Schneider}},
  \bibinfo{author}{\bibfnamefont{M.}~\bibnamefont{Gabay}},
  \bibinfo{author}{\bibfnamefont{S.}~\bibnamefont{Thiel}},
  \bibinfo{author}{\bibfnamefont{G.}~\bibnamefont{Hammerl}},
  \bibinfo{author}{\bibfnamefont{J.}~\bibnamefont{Mannhart}}, \bibnamefont{and}
  \bibinfo{author}{\bibfnamefont{J.-M.} \bibnamefont{Triscone}},
  \bibinfo{journal}{Nature} \textbf{\bibinfo{volume}{456}},
  \bibinfo{pages}{624} (\bibinfo{year}{2008}).

\bibitem[{\citenamefont{Bell et~al.}(2009)\citenamefont{Bell, Harashima,
  Kozuka, Kim, Kim, Hikita, and Hwang}}]{bell}
\bibinfo{author}{\bibfnamefont{C.}~\bibnamefont{Bell}},
  \bibinfo{author}{\bibfnamefont{S.}~\bibnamefont{Harashima}},
  \bibinfo{author}{\bibfnamefont{Y.}~\bibnamefont{Kozuka}},
  \bibinfo{author}{\bibfnamefont{M.}~\bibnamefont{Kim}},
  \bibinfo{author}{\bibfnamefont{B.~G.} \bibnamefont{Kim}},
  \bibinfo{author}{\bibfnamefont{Y.}~\bibnamefont{Hikita}}, \bibnamefont{and}
  \bibinfo{author}{\bibfnamefont{H.~Y.} \bibnamefont{Hwang}},
  \bibinfo{journal}{Phys. Rev. Lett.} \textbf{\bibinfo{volume}{103}},
  \bibinfo{pages}{226802} (\bibinfo{year}{2009}).

\bibitem[{\citenamefont{Forg et~al.}(2012)\citenamefont{Forg, Richter, and
  Mannhart}}]{forg}
\bibinfo{author}{\bibfnamefont{B.}~\bibnamefont{Forg}},
  \bibinfo{author}{\bibfnamefont{C.}~\bibnamefont{Richter}}, \bibnamefont{and}
  \bibinfo{author}{\bibfnamefont{J.}~\bibnamefont{Mannhart}},
  \bibinfo{journal}{Appl. Phys. Lett.} \textbf{\bibinfo{volume}{100}},
  \bibinfo{pages}{053506} (\bibinfo{year}{2012}).

\bibitem[{\citenamefont{Hosoda et~al.}(2013)\citenamefont{Hosoda, Hikita,
  Hwang, and Bell}}]{hosoda}
\bibinfo{author}{\bibfnamefont{M.}~\bibnamefont{Hosoda}},
  \bibinfo{author}{\bibfnamefont{Y.}~\bibnamefont{Hikita}},
  \bibinfo{author}{\bibfnamefont{H.~Y.} \bibnamefont{Hwang}}, \bibnamefont{and}
  \bibinfo{author}{\bibfnamefont{C.}~\bibnamefont{Bell}},
  \bibinfo{journal}{Appl. Phys. Lett.} \textbf{\bibinfo{volume}{103}},
  \bibinfo{pages}{103507} (\bibinfo{year}{2013}).

\bibitem[{\citenamefont{Lin et~al.}(2014)\citenamefont{Lin, Ding, Wu, Li,
  Lourembam, Shannigrahi, Wang, and Wu}}]{lin}
\bibinfo{author}{\bibfnamefont{W.-N.} \bibnamefont{Lin}},
  \bibinfo{author}{\bibfnamefont{J.-F.} \bibnamefont{Ding}},
  \bibinfo{author}{\bibfnamefont{S.-X.} \bibnamefont{Wu}},
  \bibinfo{author}{\bibfnamefont{Y.-F.} \bibnamefont{Li}},
  \bibinfo{author}{\bibfnamefont{J.}~\bibnamefont{Lourembam}},
  \bibinfo{author}{\bibfnamefont{S.}~\bibnamefont{Shannigrahi}},
  \bibinfo{author}{\bibfnamefont{S.-J.} \bibnamefont{Wang}}, \bibnamefont{and}
  \bibinfo{author}{\bibfnamefont{T.}~\bibnamefont{Wu}}, \bibinfo{journal}{Adv.
  Mat. Interf.} \textbf{\bibinfo{volume}{1}} (\bibinfo{year}{2014}), ISSN
  \bibinfo{issn}{2196-7350}.

\bibitem[{\citenamefont{Pellegrino}(2004)}]{pellegrino_thesis}
\bibinfo{author}{\bibfnamefont{L.}~\bibnamefont{Pellegrino}}, Ph.D. thesis,
  \bibinfo{school}{Università degli Studi di Genova, Dipartimento di Fisica}
  (\bibinfo{year}{2004}).

\bibitem[{\citenamefont{Pellegrino et~al.}(2002)\citenamefont{Pellegrino,
  Pallecchi, Marre, Bellingeri, and Siri}}]{pellegrino}
\bibinfo{author}{\bibfnamefont{L.}~\bibnamefont{Pellegrino}},
  \bibinfo{author}{\bibfnamefont{I.}~\bibnamefont{Pallecchi}},
  \bibinfo{author}{\bibfnamefont{D.}~\bibnamefont{Marre}},
  \bibinfo{author}{\bibfnamefont{E.}~\bibnamefont{Bellingeri}},
  \bibnamefont{and} \bibinfo{author}{\bibfnamefont{A.}~\bibnamefont{Siri}},
  \bibinfo{journal}{Appl. Phys. Lett.} \textbf{\bibinfo{volume}{81}},
  \bibinfo{pages}{3849} (\bibinfo{year}{2002}).

\bibitem[{\citenamefont{Sakudo and Unoki}(1971)}]{sakudo}
\bibinfo{author}{\bibfnamefont{T.}~\bibnamefont{Sakudo}} \bibnamefont{and}
  \bibinfo{author}{\bibfnamefont{H.}~\bibnamefont{Unoki}},
  \bibinfo{journal}{Phys. Rev. Lett.} \textbf{\bibinfo{volume}{26}},
  \bibinfo{pages}{851} (\bibinfo{year}{1971}).

\bibitem[{\citenamefont{Neville et~al.}(1972)\citenamefont{Neville, Hoeneisen,
  and Mead}}]{neville}
\bibinfo{author}{\bibfnamefont{R.}~\bibnamefont{Neville}},
  \bibinfo{author}{\bibfnamefont{B.}~\bibnamefont{Hoeneisen}},
  \bibnamefont{and} \bibinfo{author}{\bibfnamefont{C.}~\bibnamefont{Mead}},
  \bibinfo{journal}{J. Appl. Phys.} \textbf{\bibinfo{volume}{43}},
  \bibinfo{pages}{2124} (\bibinfo{year}{1972}).

\bibitem[{\citenamefont{Matthey et~al.}(2003)\citenamefont{Matthey, Gariglio,
  and Triscone}}]{matthey}
\bibinfo{author}{\bibfnamefont{D.}~\bibnamefont{Matthey}},
  \bibinfo{author}{\bibfnamefont{S.}~\bibnamefont{Gariglio}}, \bibnamefont{and}
  \bibinfo{author}{\bibfnamefont{J.-M.} \bibnamefont{Triscone}},
  \bibinfo{journal}{App. Phys. Lett.} \textbf{\bibinfo{volume}{83}},
  \bibinfo{pages}{3758} (\bibinfo{year}{2003}).

\bibitem[{\citenamefont{Stornaiuolo et~al.}(2012)\citenamefont{Stornaiuolo,
  Gariglio, Couto, Fete, Caviglia, Seyfarth, Jaccard, Morpurgo, and
  Triscone}}]{stornaiuolo}
\bibinfo{author}{\bibfnamefont{D.}~\bibnamefont{Stornaiuolo}},
  \bibinfo{author}{\bibfnamefont{S.}~\bibnamefont{Gariglio}},
  \bibinfo{author}{\bibfnamefont{N.}~\bibnamefont{Couto}},
  \bibinfo{author}{\bibfnamefont{A.}~\bibnamefont{Fete}},
  \bibinfo{author}{\bibfnamefont{A.}~\bibnamefont{Caviglia}},
  \bibinfo{author}{\bibfnamefont{G.}~\bibnamefont{Seyfarth}},
  \bibinfo{author}{\bibfnamefont{D.}~\bibnamefont{Jaccard}},
  \bibinfo{author}{\bibfnamefont{A.}~\bibnamefont{Morpurgo}}, \bibnamefont{and}
  \bibinfo{author}{\bibfnamefont{J.-M.} \bibnamefont{Triscone}},
  \bibinfo{journal}{App. Phys. Lett.} \textbf{\bibinfo{volume}{101}},
  \bibinfo{pages}{222601} (\bibinfo{year}{2012}).

\bibitem[{\citenamefont{Cancellieri et~al.}(2010)\citenamefont{Cancellieri,
  Reyren, Gariglio, Caviglia, F{\^e}te, and Triscone}}]{cancellieri}
\bibinfo{author}{\bibfnamefont{C.}~\bibnamefont{Cancellieri}},
  \bibinfo{author}{\bibfnamefont{N.}~\bibnamefont{Reyren}},
  \bibinfo{author}{\bibfnamefont{S.}~\bibnamefont{Gariglio}},
  \bibinfo{author}{\bibfnamefont{A.}~\bibnamefont{Caviglia}},
  \bibinfo{author}{\bibfnamefont{A.}~\bibnamefont{F{\^e}te}}, \bibnamefont{and}
  \bibinfo{author}{\bibfnamefont{J.-M.} \bibnamefont{Triscone}},
  \bibinfo{journal}{EPL (Europhysics Letters)} \textbf{\bibinfo{volume}{91}},
  \bibinfo{pages}{17004} (\bibinfo{year}{2010}).

\bibitem[{sig()}]{sigmavsn2D}
\bibinfo{note}{The sheet conductance regime explored in this work goes from
  0.03mS to 0.8mS and corresponds to a modulation of the carrier concentration $n_{2D}$
  from 4.8x10$^{12}$ to 1.8x10$^{13}$cm$^{-2}$.}

\bibitem[{\citenamefont{Maekawa and Fukuyama}(1981)}]{maekawa}
\bibinfo{author}{\bibfnamefont{S.}~\bibnamefont{Maekawa}} \bibnamefont{and}
  \bibinfo{author}{\bibfnamefont{H.}~\bibnamefont{Fukuyama}},
  \bibinfo{journal}{J. Phys. Soc. Jap.} \textbf{\bibinfo{volume}{50}},
  \bibinfo{pages}{2516} (\bibinfo{year}{1981}).

\bibitem[{\citenamefont{Joshua et~al.}(2012)\citenamefont{Joshua, Pecker,
  Ruhman, Altman, and Ilani}}]{joshua2012universal}
\bibinfo{author}{\bibfnamefont{A.}~\bibnamefont{Joshua}},
  \bibinfo{author}{\bibfnamefont{S.}~\bibnamefont{Pecker}},
  \bibinfo{author}{\bibfnamefont{J.}~\bibnamefont{Ruhman}},
  \bibinfo{author}{\bibfnamefont{E.}~\bibnamefont{Altman}}, \bibnamefont{and}
  \bibinfo{author}{\bibfnamefont{S.}~\bibnamefont{Ilani}},
  \bibinfo{journal}{Nature communications} \textbf{\bibinfo{volume}{3}},
  \bibinfo{pages}{1129} (\bibinfo{year}{2012}).



\bibitem[{Hall()}]{Hall}
\bibinfo{note}{Despite the unconventional measurement geometry, we observe an antisymmetric component in the magnetoresistance that we attribute to the Hall resistance of the channel. The carrier density estimated from this signal is in agreement with values observed in standard samples and its gate voltage evolution tracks the electric field sweep.} 

\bibitem[{\citenamefont{Rainer and Bergmann}(1985)}]{PhysRevB.32.3522}
\bibinfo{author}{\bibfnamefont{D.}~\bibnamefont{Rainer}} \bibnamefont{and}
  \bibinfo{author}{\bibfnamefont{G.}~\bibnamefont{Bergmann}},
  \bibinfo{journal}{Phys. Rev. B} \textbf{\bibinfo{volume}{32}},
  \bibinfo{pages}{3522} (\bibinfo{year}{1985}),

\bibitem[{Hso()}]{Hsosmall}
\bibinfo{note}{As the spin orbit characteristic field is low in this region and
  the magnetic field scale used for the estimation of $H_{i}$ and $H_{so}$ in
  Eq.\ref{MF} is narrower, the error in the determination of the crossover
  temperature is larger, as indicated by the error bars in the inset of the
  Figure.}

\bibitem[{\citenamefont{Reyren et~al.}(2009)\citenamefont{Reyren, Gariglio,
  Caviglia, Jaccard, Schneider, and Triscone}}]{reyren_apl}
\bibinfo{author}{\bibfnamefont{N.}~\bibnamefont{Reyren}},
  \bibinfo{author}{\bibfnamefont{S.}~\bibnamefont{Gariglio}},
  \bibinfo{author}{\bibfnamefont{A.~D.} \bibnamefont{Caviglia}},
  \bibinfo{author}{\bibfnamefont{D.}~\bibnamefont{Jaccard}},
  \bibinfo{author}{\bibfnamefont{T.}~\bibnamefont{Schneider}},
  \bibnamefont{and} \bibinfo{author}{\bibfnamefont{J.-M.}
  \bibnamefont{Triscone}}, \bibinfo{journal}{Appl. Phys. Lett.}
  \textbf{\bibinfo{volume}{94}}, \bibinfo{eid}{112506} (\bibinfo{year}{2009}).

\bibitem[{\citenamefont{Clogston}(1962)}]{clogston}
\bibinfo{author}{\bibfnamefont{A.~M.} \bibnamefont{Clogston}},
  \bibinfo{journal}{Phys. Rev. Lett.} \textbf{\bibinfo{volume}{9}},
  \bibinfo{pages}{266} (\bibinfo{year}{1962}).

\bibitem[{\citenamefont{Wu et~al.}(2006)\citenamefont{Wu, Adams, Yang, and
  McCarley}}]{wu}
\bibinfo{author}{\bibfnamefont{X.~S.} \bibnamefont{Wu}},
  \bibinfo{author}{\bibfnamefont{P.~W.} \bibnamefont{Adams}},
  \bibinfo{author}{\bibfnamefont{Y.}~\bibnamefont{Yang}}, \bibnamefont{and}
  \bibinfo{author}{\bibfnamefont{R.~L.} \bibnamefont{McCarley}},
  \bibinfo{journal}{Phys. Rev. Lett.} \textbf{\bibinfo{volume}{96}},
  \bibinfo{pages}{127002} (\bibinfo{year}{2006}).

\bibitem[{\citenamefont{Salluzzo et~al.}(2009)\citenamefont{Salluzzo, Cezar,
  Brookes, Bisogni, De~Luca, Richter, Thiel, Mannhart, Huijben, Brinkman
  et~al.}}]{salluzzo}
\bibinfo{author}{\bibfnamefont{M.}~\bibnamefont{Salluzzo}},
  \bibinfo{author}{\bibfnamefont{J.}~\bibnamefont{Cezar}},
  \bibinfo{author}{\bibfnamefont{N.}~\bibnamefont{Brookes}},
  \bibinfo{author}{\bibfnamefont{V.}~\bibnamefont{Bisogni}},
  \bibinfo{author}{\bibfnamefont{G.}~\bibnamefont{De~Luca}},
  \bibinfo{author}{\bibfnamefont{C.}~\bibnamefont{Richter}},
  \bibinfo{author}{\bibfnamefont{S.}~\bibnamefont{Thiel}},
  \bibinfo{author}{\bibfnamefont{J.}~\bibnamefont{Mannhart}},
  \bibinfo{author}{\bibfnamefont{M.}~\bibnamefont{Huijben}},
  \bibinfo{author}{\bibfnamefont{A.}~\bibnamefont{Brinkman}},
  \bibnamefont{et~al.}, \bibinfo{journal}{Phys. Rev. Lett.}
  \textbf{\bibinfo{volume}{102}}, \bibinfo{pages}{166804}
  (\bibinfo{year}{2009}).

\bibitem[{\citenamefont{BenShalom et~al.}(2010{\natexlab{b}})\citenamefont{BenShalom,
  Ron, Palevski, and Dagan}}]{shalom2010shubnikov}
\bibinfo{author}{\bibfnamefont{M.}~\bibnamefont{BenShalom}},
  \bibinfo{author}{\bibfnamefont{A.}~\bibnamefont{Ron}},
  \bibinfo{author}{\bibfnamefont{A.}~\bibnamefont{Palevski}}, \bibnamefont{and}
  \bibinfo{author}{\bibfnamefont{Y.}~\bibnamefont{Dagan}},
  \bibinfo{journal}{Physical review letters} \textbf{\bibinfo{volume}{105}},
  \bibinfo{pages}{206401} (\bibinfo{year}{2010}{\natexlab{b}}).

\bibitem[{\citenamefont{Caviglia
  et~al.}(2010{\natexlab{b}})\citenamefont{Caviglia, Gariglio, Cancellieri,
  Sac{\'e}p{\'e}, Fete, Reyren, Gabay, Morpurgo, and
  Triscone}}]{caviglia2010two}
\bibinfo{author}{\bibfnamefont{A.D.}~\bibnamefont{Caviglia}},
  \bibinfo{author}{\bibfnamefont{S.}~\bibnamefont{Gariglio}},
  \bibinfo{author}{\bibfnamefont{C.}~\bibnamefont{Cancellieri}},
  \bibinfo{author}{\bibfnamefont{B.}~\bibnamefont{Sac{\'e}p{\'e}}},
  \bibinfo{author}{\bibfnamefont{A.}~\bibnamefont{Fete}},
  \bibinfo{author}{\bibfnamefont{N.}~\bibnamefont{Reyren}},
  \bibinfo{author}{\bibfnamefont{M.}~\bibnamefont{Gabay}},
  \bibinfo{author}{\bibfnamefont{A.F.}~\bibnamefont{Morpurgo}}, \bibnamefont{and}
  \bibinfo{author}{\bibfnamefont{J.-M.} \bibnamefont{Triscone}},
  \bibinfo{journal}{Physical review letters} \textbf{\bibinfo{volume}{105}},
  \bibinfo{pages}{236802} (\bibinfo{year}{2010}{\natexlab{b}}).

\end{thebibliography}
\end{document}